\documentclass[aps,prmaterials,notitlepage,superscriptaddress,reprint]{revtex4-2}
\usepackage[final]{graphicx}
\usepackage{natbib}
\usepackage{hyperref}
\usepackage{amssymb}
\usepackage{amsmath}
\usepackage{mathrsfs}
\usepackage{xcolor}
\usepackage{bm}

\begin{document}
\title{Strong atomic reconstruction in twisted bilayers of highly flexible InSe: Machine-Learned Interatomic Potential and continuum model approaches}

\author{S. J. Magorrian}
\affiliation{Department of Physics, University of Warwick, Coventry, CV4 7AL, United Kingdom}

\author{A. Siddiqui}
\affiliation{Department of Physics, University of Warwick, Coventry, CV4 7AL, United Kingdom}

\author{N. D. M. Hine}
\email{n.d.m.hine@warwick.ac.uk}
\affiliation{Department of Physics, University of Warwick, Coventry, CV4 7AL, United Kingdom}


\begin{abstract}
The relaxation of atomic positions to their optimal structural arrangement is crucial for understanding the emergence of new physical behavior in long scale superstructures in twisted bilayers of two-dimensional materials. The amount of deviation from a rigid moiré structure will depend on the elastic properties of the constituent monolayers which for the twisted bilayer - the more flexible the monolayers are, the lower the energy required to deform the layers to maximize the areas with an energetically optimal interlayer arrangement of atoms. We investigate this atomic reconstruction for twisted bilayers of highly flexible InSe. Results using two methods are demonstrated - first we train a machine-learned interatomic potential (MLIP) to enable fully atomistic relaxations of small-twist-angle large-length-scale moiré supercells while retaining density functional theory (DFT) level accuracy. We find substantial out-of-plane corrugation and in-plane domain formation for a wide range of twist angles and moiré length scales. We then adapt an existing continuum approach and show that it can reproduce some, but not all, features of the fully atomistic calculations.
\end{abstract}

\maketitle

\section{Introduction}
Stacking homo- or hetero-bilayers of two-dimensional (2D) materials with a twist and/or lattice mismatch between the layers leads to long-range periodic modulations of atomic structure, known as moiré patterns. A variety of moiré-scale physical phenomena arise as a result of the modulation of electronic and mechanical properties, which are not necessarily apparent at the shorter length scale of primitive aligned bilayers.
Extensive research has uncovered varied new physics in twisted bilayers of graphene\cite{tbg_2018_1, tbg_2018_2, Yankowitz_2019, Kerelsky_2019, Choi_2019}, and of the transition metal dichalcogenides (TMDs)\cite{Yu_2017, Alexeev_2019, PhysRevB.99.125424, Seyler_2019, Tran_2019, Jin_2019, Wang_2020, PhysRevLett.124.217403}. In both cases, behaviour diverging greatly from monolayers and aligned bilayers is obtained as a result of flat bands and the resulting van Hove singularities that occur in the density of states\cite{Bistritzer_2011, PhysRevLett.121.026402, PhysRevLett.122.086402}. The relative importance of lattice relaxations and electron-electron correlation in driving the emergence of flat bands has been extensively debated, and it is clear that both effects play a role\cite{Molino_2023, Li_2024}.

%

It is therefore crucial to fully characterize the structure and arrangement of atoms within a moiré supercell of a twisted bilayer. 
A moiré pattern formed in a rigid twisted bilayer will feature a continuously varying relative arrangement of atoms in the two layers.
Where it becomes energetically favorable, the two layers can deform to maximise areas of the supercell with lower-energy atomic arrangements, at the expense of areas with less-favorable atomic stackings.
This atomic reconstruction can transform the structure of a twisted bilayer, from a smoothly varying moiré arrangement to a pattern of energetically-favorable domains separated by a network of domain-wall dislocations and nodes\cite{PhysRevB.98.224102, PhysRevLett.124.206101}.
This transformation has interlayer effects, in terms of the interlayer atomic arrangements and hybridization of electronic states, as well as intralayer effects, due to the highly non-uniform strain patterns within the layers formed on reconstruction.
Effects of reconstruction and associated domain formation have been observed in graphene\cite{Yoo_2019}, hexagonal boron nitride (hBN)\cite{PhysRevB.103.125427, Xian_2019, Yasuda_2021, Woods_2021}, and the TMDs\cite{Sung_2020, Weston_2020, Li_2021, Tilak_2022, Weston_2022, Ko_2023, Van_Winkle_2023, LuicanMayer_2023, Foutty_2024}.

Beyond these materials, the III-VI layered post-transition metal chalcogenide (PTMC) InSe has properties of great importance in the development of 2D materials for the exploration of new physical phenomena and technological applications.
%
The conduction band has a light electron effective mass, leading to the achievement of mobilities exceeding 10$^4$~cm$^2$V$^{-1}$s$^{-1}$ in few-layer films, and the observation of fully developed quantum Hall behavior\cite{Bandurin_2016}.
This has lead to InSe being demonstrated as a basis for 2D material-based field-effect transistors\cite{Feng_2014, Ho_2017, Hu_2022}
The valence band has holes with a somewhat heavier effective mass, which on going to the thinnest films develops a slightly offset maximum\cite{Mudd_2016, Kibirev_2018, Hamer_2019, Pasquale_2023}. 
This gives a direct to slightly indirect transition in the character of the gap on going from bulk to few-layers, in contrast to the indirect-direct transition in the TMDs.
This nearly flat `Mexican hat' dispersion in few-layer InSe features a Van Hove singularity near the band edge, with the high density of states leading to predictions of strongly-correlated behavior in p-doped films\cite{PhysRevB.89.205416, PhysRevLett.123.176401, PhysRevB.103.035411, Stepanov_2022}.
The size of the band gap varies substantially with the number of layers, going from 1.35~eV\cite{PhysRevB.17.4718} in the bulk up to $\sim$2.8~eV\cite{PhysRevB.94.245431} in the monolayer, with the resulting optical activity covering the visible spectrum\cite{Bandurin_2016, Hamer_2019, Mudd_2013, Lei_2014, Zheng_2017}. 
Optical activity in the infrared can be attained by exploitation of layer-number-dependent well-formed semiconductor quantum well-like subbands in few layer films\cite{PhysRevB.97.165304, Zultak_2020, Shcherbakov_2024}.

This strong layer-number dependence in optical and electronic properties of InSe indicates sensitivity of band-edge electrons/holes to interlayer hybridization, and a likely sensitivity to changes in stacking in twisted bilayers\cite{Kang_2020}. 
Together with the high flexibility of InSe\cite{PhysRevB.95.115409, Zhao2019, Li2019}, which reduces the energy cost of reconstruction, 
this motivates the study of twisted bilayer InSe and suggest it will be a good candidate for emergent new physics resulting from domain formation.

In modelling such twisted bilayer systems, due to the high computational cost of Density Functional Theory (DFT) calculations calculations on moiré-scale systems, is common to use continuum models based on elasticity theory and parameterised using mechanical properties calculated via DFT.
However, the accuracy of continuum models is not well-characterised, particularly in the regime of intermediate twist.
Here, we investigate the structural relaxation and reconstruction of twisted bilayer InSe using a fully-atomistic approach of accuracy equivalent to DFT, by training a machine-learned interatomic potential (MLIP) utilising the MACE equivariant neural network\cite{Batatia2022mace}.

We find that substantial atomic reconstruction of the twisted bilayer is present for a wide range of twist angles due to the high flexibility of the InSe layers. 
For twist angles $< 5^{\circ}$ a clear pattern of low stacking-energy domains separated by a network of domain walls and nodes emerges.
We also adapt the continuum model for relaxation of twisted TMD bilayers in Ref.~\onlinecite{PhysRevLett.124.206101} to the InSe case, and compare its results for the smaller twist angles to those predicted using an atomistic approach.
We find that while the model reproduces well the domains for twist angles near 0$^{\circ}$, there are differences in the projected interlayer distances, which we show results from the way the continuum approach neglects the energy costs of out-of-plane distortions.
In contrast, for twist angles near $60^{\circ}$, the continuum model not only fails to reproduce interlayer distances but also fails to fully reproduce the domain shapes found by atomistic relaxation.

This paper is structured as follows: In Section~\ref{sec:MACE} we present results of atomistic relaxations of twisted InSe bilayers via the use of a machine-learned interatomic potential (MLIP). 
In Section~\ref{sec:continuum} we present a continuum model parametrised from the same DFT results, and discuss the differences between the two approaches.
We conclude our findings in Section~\ref{sec:conclusions}.
\section{Machine-learned interatomic potential}
\label{sec:MACE}
\subsection{Training}
Density functional theory (DFT) calculations, on which the MLIP and continuum models in this paper are based, are performed using the QuantumESPRESSO code\cite{Giannozzi2009, Giannozzi2017, Giannozzi2020}. 
We employ the OPTB88 van der Waals functional\cite{PhysRevLett.115.136402,PhysRevB.76.125112,Berland2015,Langreth2009,Sabatini2012}, and potentials from the SSSP library\cite{Prandini2018} - specifically the In PBE pseudopotential from PSlibrary v0.3.1\cite{arxiv.1404.3015} and the Se PBE pseudopotential from the GBRV dataset\cite{Garrity2014}. 
We use equivalent, well-converged k-point grids across all system sizes: 12$\times$12$\times$1 for the primitive monolayer and bilayer cells, 2$\times$2$\times$1  for the 6$\times$6 monolayer supercells, and 5$\times$5$\times$1 and 3$\times$3$\times$1 grids for the 21.79$^{\circ}$ and 13.17$^{\circ}$ cells, respectively.
The kinetic energy and charge density cutoffs were 80~Ry and 720~Ry respectively, providing accurate forces and stresses.

\begin{figure}
    \centering
    \includegraphics[width = \linewidth]{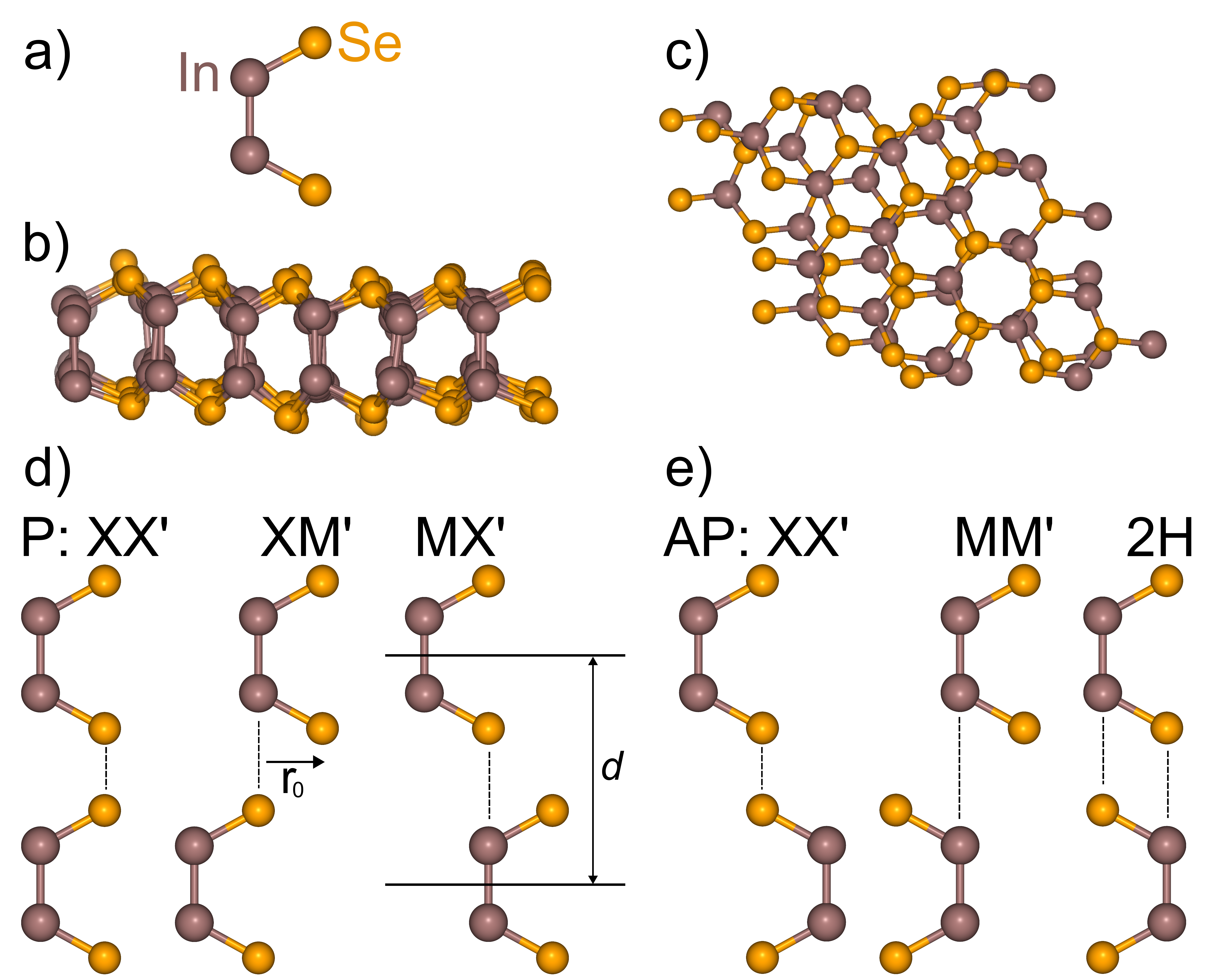}
    \caption{Schematics of InSe configurations selected from the MLIP training dataset. a) Primitive monolayer cells. b) 6$\times$6 monolayer cells with atoms displaced via MD. c) Large angle small moiré cell twisted bilayers. d) and e) aligned primitive bilayers, with high-symmetry stackings shown.}
    \label{fig:structures}
\end{figure}
The training set consists of DFT energies, forces and stresses calculated for several sets of configurations, designed to ensure the potential correctly reproduces DFT results for both intralayer and interlayer interactions. To these, we add the energies of isolated In and Se atoms, to anchor the energy contribution associated with each species.

To learn intralayer interactions and energies, we include calculations on primitive cells of the InSe monolayer and 6$\times$6 supercells. In the former case, we model both a fully relaxed unstrained cell, and cells subject to uniaxial, biaxial and shear strains up to 4\% (25 configurations). 
For the 6$\times$6 supercells, a range of possible intralayer relative atomic positions is sampled by displacing the atoms by running multiple Molecular Dynamics (MD) simulations on the pristine monolayers to generate equilibrated configurations, at a range of 7 temperatures from 50~K to 600~K, and 9 strain configurations (uniaxial and biaxial) from -2\% to 2\%.
The temperature range extends to a point below the melting point of InSe ($\sim 900$~K)\cite{Klemm_1934, Celustka_1974}, while the strain range extends well beyond the nonuniform internal strain range expected within low-twist angle bilayer systems, which gives the resulting MLIP the ability to explore all intralayer relative atomic displacements and in-plane strains which may realistically be encountered in the relaxation of a twisted bilayer.
%
These initial MD simulations make use of the MACE-mp0 foundation model, a general-purpose pre-trained ML model\cite{batatia2023foundation} trained on crystal structures from across the whole periodic table.
Prior to using MACE-mp0 for this purpose, we confirmed that it accurately reproduced the structure of monolayer InSe, but that it does not reproduce stacking energetics of aligned bilayers, nor the vibrational properties of the resulting models, with sufficient accuracy, meriting the training of a single-purpose MACE MLIP for the current work.

MD simulations using MACE-mp0 are run for 5000~fs with a timestep of 5~fs for each combination of temperature and strain, regulated by a Langevin thermostat employing a damping of 0.1~fs$^{-1}$.
Energies and forces are then calculated using DFT for the atomic positions of the final frame of each trajectory, for a total of 63 thermally-disordered configurations of the 6$\times$6 monolayer.

To ensure the potential learns interlayer interactions, singlepoint energy and force calculations are performed for rigid primitive bilayer cells constructed from relaxed monolayers. 
The layers are displaced relative to each other by in-plane vectors $\mathbf{r}_0$ drawn from the irreducible portion of a 6$\times$6 grid covering the primitive monolayer cell, and out-of-plane by distances $d$ spanning from 0.82~nm to 0.94~nm in steps of 0.01~nm. 
This is done with the layers aligned both parallel (P-stacking) and antiparallel (AP-stacking) to each other.
The P case gives fewer irreducible in-plane shifts since displacing one layer by +/- $\mathbf{r}_0$ gives a pair of equivalent twins. In total this gives 572 primitive bilayer configurations.
Finally, we include snapshots for the two large-angle twisted bilayers easilytractable via DFT. These are for twist angles 21.79$^{\circ}$ and 13.17$^{\circ}$, and configurations are included for both rigid twisted bilayers with interlayer distance 0.875~nm, and the same systems following relaxation.
In total this gives 666 configurations in the training set.
Examples of structures taken from each category in the training set are illustrated in Fig.~\ref{fig:structures}.

The model is then trained using the MACE equivariant neural network\cite{Batatia2022mace}. A 90\%-5\%-5\% train-test-validate split was employed on the training set. 
We use two NN layers and a cutoff radius of 8~\AA~for the atomic environment. This cutoff radius, greater than the usual default of 5~\AA~in MACE, ensures that interlayer interactions can be reproduced accurately, particularly the dependence on stacking. The batch size was set to 10 and the model was trained for a total of 1200 epochs. 
The ratio of energy, force and stress weights was 1:10:1 for the first 200 epochs, then Stochastic Weight Averaging \cite{Izmailov_SWA_2018} is activated and the weights changed to 100:10:1 thereafter, to improve final convergence of the energy. 
The loss function was minimized using the AMSGrad variant of the Adam\cite{Kingma2014} optimizer with a learning rate of 0.01.
RMS errors on the energies for both the training and validation proportions of the set are 0.1~meV/atom. For the forces the RMS errors are 6.4~meV\AA$^{-1}$ and 7.1~meV\AA$^{-1}$ for the training and validation sets, respectively.
The force errors are found predominantly in the higher-temperature MD-displaced 6$\times$6 monolayer cells, where the atoms are most displaced from their equilibrium positions.

\begin{figure}
    \centering
    \includegraphics[width = \linewidth]{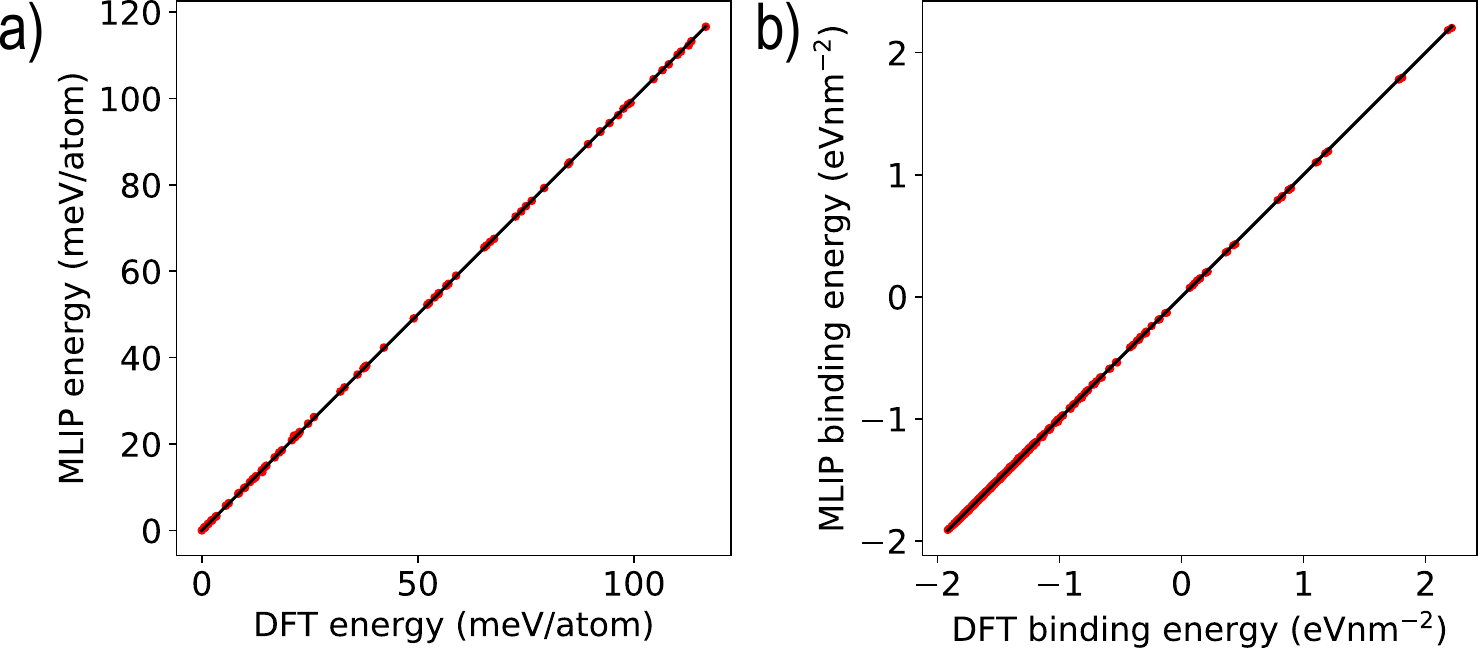}
    \caption{a) Scatter plot of MLIP energies vs DFT energies for unstrained and strained primitive and 6$\times$6 InSe monolayers, the line indicating where DFT energy and MLIP energy are equal. Energies relative to the unstrained primitive monolayer. b) Scatter plot of MLIP vs. DFT binding energies for all aligned primitive bilayers included in training set.}
    \label{fig:mace_scatter}
\end{figure}

Fig.~\ref{fig:mace_scatter}(a) shows a scatter plot comparing the DFT energies of all monolayer configurations included in the training and validation set with those predicted by the MLIP, showing good agreement between the trained model and DFT. 

\begin{figure}
    \centering
    \includegraphics[width = \linewidth]{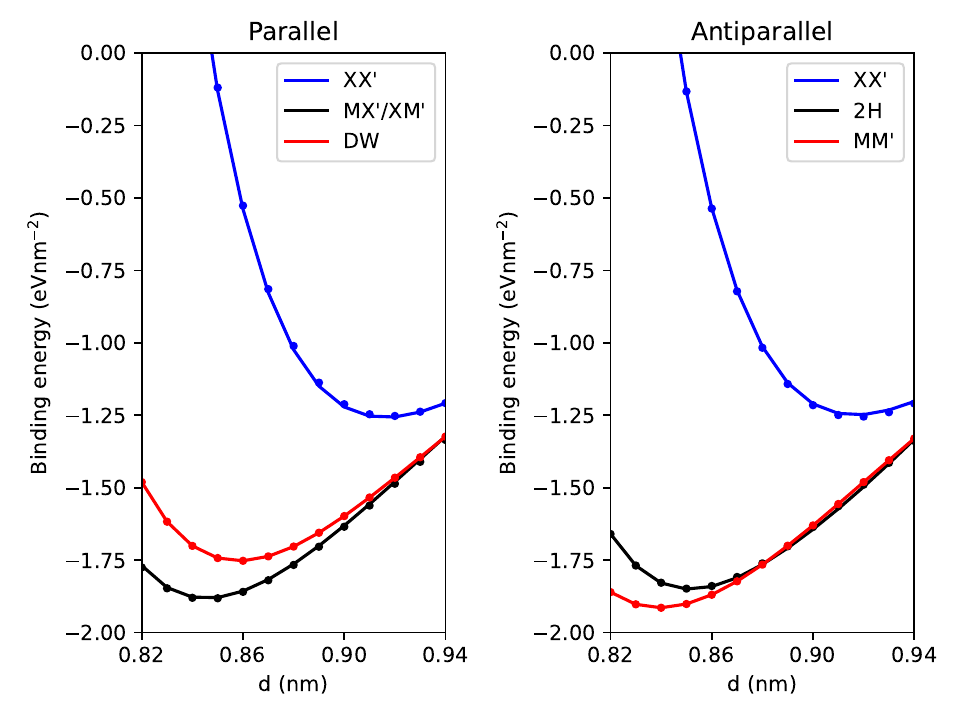}
    \caption{Comparison between DFT data (dots) and energies from MLIP(lines), for selected $\mathbf{r}_0$ stacking configurations as a function of interlayer distance $d$. The DW (domain wall) stacking in the parallel case lies midway along the shortest path between the MX$'$ and XM$'$ twins.}
    \label{fig:bilayer_mace_dft}
\end{figure}

A scatter plot comparing DFT energies to those calculated using the MLIP for all bilayer configurations in the training and validation set can be found in Fig.~\ref{fig:mace_scatter}(b).
In Fig.~\ref{fig:bilayer_mace_dft}, we show for the high-symmetry bilayer atomic registries the dependence of the interlayer binding energy on the distance $d$ between the mean planes of the two layers. 
These show a very good agreement between DFT and MLIP interlayer binding energies.
The dominant factor determining the dependence of the binding energy on the atomic registry is the relative positions of the Se atoms - in both the parallel and antiparallel cases the XX' stacking has a higher minimum energy (i.e. a weaker binding) at a greater interlayer distance than the other stackings.
In the parallel case, stackings related by an inversion of the in-plane relative shift of the layers ($\mathbf{r}_0$), e.g. MX$'$ and XM$'$, have the same energy.
In the antiparallel case, this symmetry is broken, and the related stackings MM' and 2H now have slightly different binding energies and are also separated in their optimal interlayer distance.
In contrast to the dominant 2H polytype found for the TMDs\cite{Bronsema1986, Schutte1987}, for InSe we find that the MM$'$ is the lower energy stacking of the two, with a smaller optimal interlayer distance. 
The calculations presented here also predict that the minimum energy of the antiparallel MM$'$ stacking order is lower than that of the parallel MX$'$/XM$'$ twin stackings.
DFT calculations for bulk crystals, using a variety of vdW correction methods, show similar energetic ordering\cite{PhysRevB.103.094118}.

\subsection{Strong atomic reconstruction from large twist angles}
Having obtained a trained and validated potential, we proceed to geometry relax models of the twisted bilayer. We present results for a range of P-stacked and AP-stacked twisted bilayer models, chosen to have minimal-sized commensurate cells at a range of specific angles. This approach, described in Ref.~\cite{dosSantos_2007}, generates bilayer supercells with relative twist angle $\theta$ given by
\begin{equation}
    \cos\theta = \frac{3i^2+3i+\frac{1}{2}}{3i^2+3i+1} \qquad i=0,1,2,\dots
\end{equation}
whose moiré lattice period $L_M$ is given by
\begin{equation}
L_M = \sqrt{3i^2+3i+1}a_0
\end{equation}
where $a_0$ is the primitive cell size. The supercell indexed $i$ contains $N=8(3i^2+3i+1)$ atoms. We present results for $i=1,2,5,10,20,40$ to illustrate trends as $\theta$ goes from near 30$^\circ$ to near 0$^\circ$ for P-stacking, and from near 30$^\circ$ to near 60$^\circ$ for AP-stacking. For P-stacking, $\theta < 1^\circ$ for the largest value, $i=40$ by which point $L_M=285$~\AA.

For twist angles $> 2.5^{\circ}$ (up to 3752 atoms), relaxation was carried out using the optimization module of the Atomic Simulation Environment\cite{HjorthLarsen2017}. For smaller twist angles, we exported the compiled MACE potential and used LAMMPS\cite{Thompson2022} for large-scale geometry optimisations.
We considered the geometry converged when the forces on all atoms had a maximum magnitude component $<$0.25~meV/Å. 

Atomic reconstruction in twisted bilayers is governed by the competition between (i) the energy to be gained by maximizing the areas of lowest energy (i.e. strongest binding) relative atomic registries between the two layers and (ii) the energy cost of deforming the layers to do so.
This competition sets a length scale for the size of domain walls and nodes, and reconstruction becomes apparent once the twist angle becomes small enough that these features can fit inside a moiré supercell of the twisted bilayer.

While the energy differences between the highest- and lowest-energy stackings (shown in Fig.~\ref{fig:bilayer_mace_dft} are similar to those found for the TMDs\cite{PhysRevLett.124.206101}, the elastic properties, and hence the energy cost of deforming the InSe layers, are quite different.
Experimental determinations of Young's modulus for few-layer InSe show a wide range, from as low as 23(5)~GPa\cite{Zhao2019} up to around 100GPa\cite{Li2019}.
Approximating the effective monolayer thickness of InSe as $\sim 0.85$~nm, theoretical values extracted from the DFT calculations in this study give a Young's modulus $\sim 50$~GPa. 
All these values are smaller than those found for the TMDs ($\sim 150-350$~GPa) and substantially smaller than those of graphene - a useful summary table comparing Young's moduli for a range of 2D materials can be found in Ref. \onlinecite{Zhao2019}. 
The consequence of the lower Young's modulus is that we can expect the resulting lower energy cost of deforming the layers to lead to reconstruction appearing at larger twist angles than for TMDs.

\begin{figure*}
    \centering
    \includegraphics[width = \linewidth]{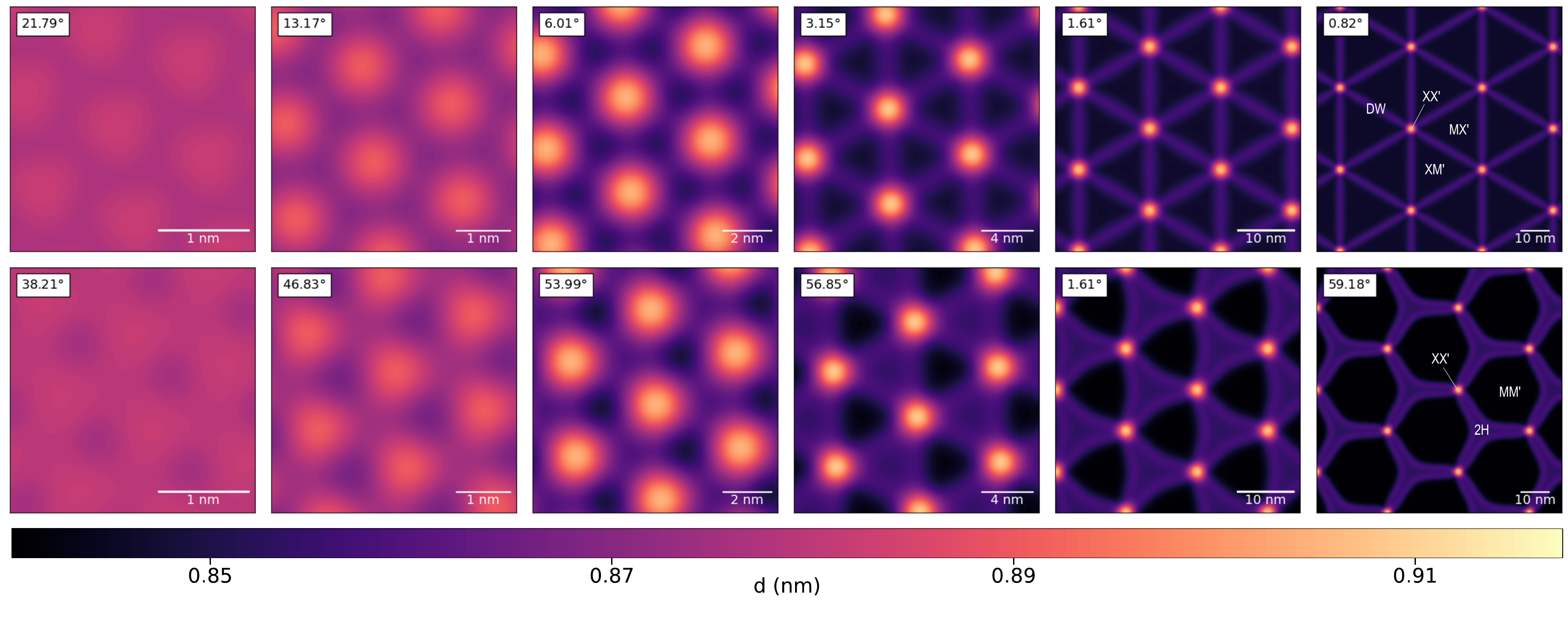}
    \caption{Maps of interpolated interlayer distances (taken as the difference between the mean of the out-of-plane coordinates of the two In sublayers in each layer) for MLIP-relaxed twisted InSe bilayers at selected angles indicated. Annotation indicates locations of high-symmetry stackings.}
    \label{fig:mace_ild_maps}
\end{figure*}

The dependence of adhesion energy on both in-plane shift and interlayer distance (Fig.~\ref{fig:bilayer_mace_dft}) allows visualization of the effects of relaxation through in-plane interpolation of the vertical distance between the layers. 
In Fig.~\ref{fig:mace_ild_maps}, we show images of the reconstruction of twisted InSe bilayers for a range of twist angles. 
We find a departure from a rigid twisted bilayer for all twist angles.
Corrugation in the interlayer distance, apparent at all angles, becomes greater for smaller twist angles/larger moir{\'e} length scales.
A pattern of triangular domains of the energetically favorable MX$'$/XM$'$ stackings separated by a network of domain walls and XX$'$ stacked nodes is already apparent at $\theta \sim 6^{\circ}$, and fully-formed by $\theta \sim 3^{\circ}$. 
These domains persist in similar form down to the smallest twist angles considered.

The reconstruction of AP-bilayers (twist angle near $60^{\circ}$) has two distinct phases, depending on twist angle.
At large twist angles, the reconstructed moir{\'e} patterns of AP-bilayers look very similar to the P-stacked case, with XM$'$/MX$'$ stacked regions replaced by MM$'$ and 2H stacking. 
This happens since the dominant effect is the minimization of the area of the high-energy XX$'$ stacked regions. 
For smaller angles, the asymmetry of the interlayer binding energy under inversion of $\mathbf{r}_0$ becomes apparent, with the slightly favorable MM$'$ stacked domains growing at the expense of the 2H stacked regions.

\begin{figure}
    \centering
    \includegraphics[width = \linewidth]{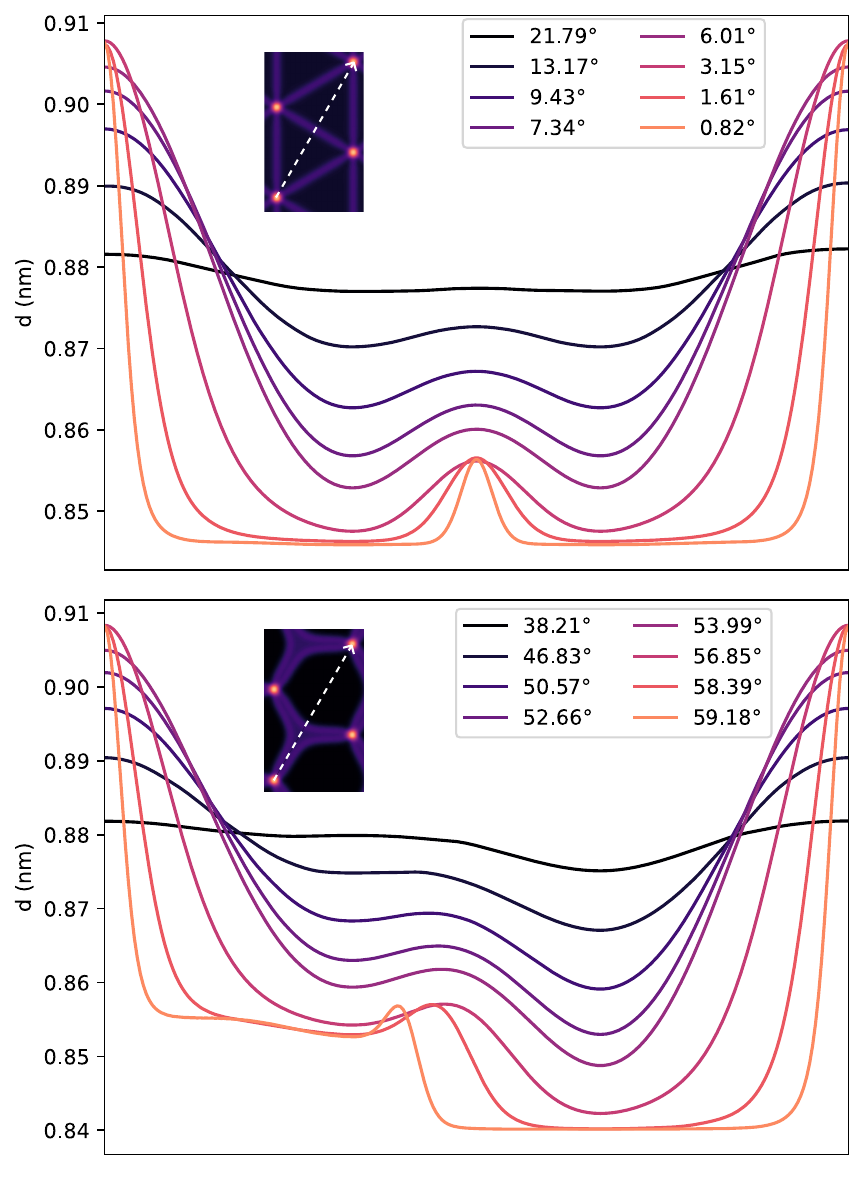}
    \caption{Variation of interlayer distance along a path of length $\sqrt{3}L_M$ across the moiré supercell for MLIP-relaxed twisted InSe bilayers, for twist angles mapped in Fig.~\ref{fig:mace_ild_maps}.}
    \label{fig:mace_ild_lines}
\end{figure}

\section{Continuum model}
\label{sec:continuum}
The procedure for constructing a continuum model to describe reconstruction in moiré superlattices of twisted bilayer InSe broadly follows that used for the transition metal dichalcogenides (TMDs) which is described in Ref. \onlinecite{PhysRevLett.124.206101}.
Nevertheless, InSe exhibits differences in the behavior of interlayer adhesion as a function of local interlayer distance ($d$) and in-plane shift ($\mathbf{r}_0$).
These differences give rise to a requirement for some small modifications to the procedure used previously for the TMDs, which we set out below.
\subsection{Interpolation of DFT-calculated adhesion energies for aligned bilayers}
The DFT calculations of adhesion energy as a function of interlayer distance $d$ and of in-plane shift $\mathbf{r}_0$ are interpolated using the expression (adapted from Ref. \onlinecite{PhysRevLett.124.206101}):
\begin{align}\label{eq:adhesion}
    W(d,\mathbf{r}_0) = &\sum_{n=1}^{3}\frac{C_n}{(d-0.2)^{4n}}\\
    &+\sum_{n=1}^{3}\sum_{j=1}^3\bigg(A_ne^{-\alpha_n (d-d_0)}\cos(\mathbf{G}^n_j\cdot \mathbf{r}_0)\nonumber\\
    &+(B_ne^{-\beta_n (d-d_0)}+B_n'e^{-\beta_n' (d-d_0)})\nonumber\\
    &\quad\times\sin(\mathbf{G}^n_j\cdot \mathbf{r}_0+\phi)\bigg).\nonumber
\end{align}
Here, $\mathbf{G}^n_j$ is the $j$th vector in the $n$th star of the InSe reciprocal lattice, with terms up to the 3rd star included. $\phi=0$ for AP-InSe and $\phi=\pi/2$ for P-InSe. 
A shift to the interlayer distance in the $\mathbf{r}_0$-independent part is applied to ensure that the fit has a negative tail at large $d$. It is found that, in contrast to the previous work studying the TMDs, the $d$-dependence of the $\mathbf{r}_0$ antisymmetric part cannot be fitted well with a single exponential, and two exponential terms are used to ensure a good fit. 

\begin{figure}
    \centering
    \includegraphics[width = \linewidth]{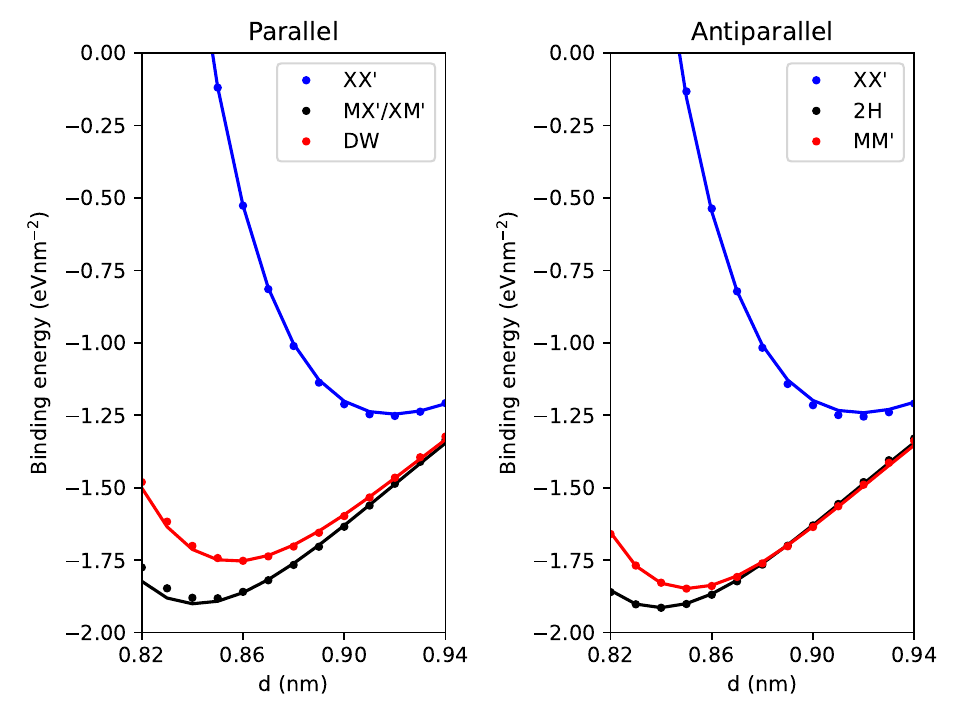}
    \caption{Comparison between DFT data (dots) and continuum model fit (lines), Eq. \eqref{eq:adhesion}, for selected $\mathbf{r}_0$ stacking configurations as a function of interlayer distance $d$.}
    \label{fig:bilayer_cont_dft}
\end{figure}

To determine the parameters in Eq. (\ref{eq:adhesion}), we calculate DFT energies in the irreducible portion of a 24$\times$24$\times$1 grid in $\mathbf{r}_0$ space, for the same 0.82~nm -- 0.94~nm range of interlayer distances $d$ as used above in the MLIP training.
At each $d$, the average DFT energy and Fourier components of its dependence on $\mathbf{r}_0$ are determined by a numerical Fourier transform of the DFT data for AP-InSe. 
A fit of the dependence on $d$ of the Fourier components is then performed to determine the constants required in Eq.~(\ref{eq:adhesion}). 
The resulting parameters are summarised in Table~\ref{tab:parameters}. 
We compare the energy curves generated by  Eq.~(\ref{eq:adhesion}) for a selected set of $\mathbf{r}_0$ configurations against the corresponding DFT points in Fig.~\ref{fig:bilayer_cont_dft}.
\begin{table}
\caption{Parameters in Eq. (\ref{eq:adhesion}), from fitting ($C_n$) and numerical Fourier transform (other parameters) of DFT data. 
\label{tab:parameters}
}
\begin{tabular}{cccc}
 \hline
 \hline
 $C_1$ & $-0.250$ $\mathrm{eV\cdot nm^{2}}$ & $d_0$ & 0.87 nm \\
 $C_2$ & $-0.106$ $\mathrm{eV\cdot nm^{6}}$& & \\
 $C_3$ & $0.0192$ $\mathrm{eV\cdot nm^{10}}$&  & \\
 \hline
 $A_1$ & 0.108 eV/nm$^2$&$\alpha_1$ & 27.4 nm$^{-1}$ \\
 $A_2$ & 5.42$\times 10^{-3}$ eV/nm$^2$&$\alpha_2$ & 34.2 nm$^{-1}$\\
 $A_3$ & 1.67$\times 10^{-3}$ eV/nm$^2$&$\alpha_3$ & 37.2 nm$^{-1}$ \\
 \hline
 $B_1$ & $-3.24\times 10^{-3}$ eV/nm$^2$&$\beta_1$ & 13.1 nm$^{-1}$ \\ 
 $B_1'$ & 5.26$\times 10^{-3}$ eV/nm$^2$&$\beta_1'$ & 34.4 nm$^{-1}$ \\ 
 $B_3$ & $-5.15\times 10^{-5}$ eV/nm$^2$&$\beta_3$ & 19.8 nm$^{-1}$ \\
 $B_3'$ & $2.36\times 10^{-4}$ eV/nm$^2$&$\beta_3'$ & 42.2 nm$^{-1}$ \\ 
 \hline
\end{tabular}
\end{table}

\subsection{Description of reconstruction in twisted bilayers}
To obtain a description of lattice reconstruction in twisted bilayer InSe, we aim to minimize the sum of the adhesion energy $W$ and the local elastic energy cost, $U(\mathbf{r})$, due to the strain arising from the reconstruction of the lattices
\begin{equation}
    \label{eq:competition}
    \mathcal{E} = \int d^2\mathbf{r} \left[U(\mathbf{r}) + W(d(\mathbf{r}), \mathbf{r}_0(\mathbf{r}))\right].
\end{equation}
$U(\mathbf{r})$ is given by
\begin{equation}
\label{eq:strain_cost}
    U(\mathbf{r}) = \lambda_l (u_{ii}(\mathbf{r}))^2 + 2\mu_l u_{ij}^2(\mathbf{r})
\end{equation}
where $u_{ij} = \frac{1}{2}(\partial_ju_i+\partial_iu_j)$ are the strain tensors resulting from local atomic displacement, $\mathbf{u(r)}$, in the top InSe layer. $\mathbf{u(r)}$ is expanded in a Fourier series over the reciprocal lattice vectors, $\mathbf{g}_n$, of the moir{\'e} superlattice,
\begin{equation}
\label{eq:ur_expansion}
     \mathbf{u(r)} = \sum_{\mathbf{g}_n} \mathbf{u}_{\mathbf{g}_n}e^{i\mathbf{g}_n\cdot \mathbf{r}}.  
\end{equation}
with the local lateral offset in the twisted bilayer then being approximated as
\begin{equation}
    \mathbf{r}_0(\mathbf{r}) \simeq \theta\hat{z}\times\mathbf{r} + 2\mathbf{u}(\mathbf{r}).
\end{equation}
Here, since we consider homobilayers of InSe, we have set the atomic displacements in the top and bottom layers to be equal and opposite to each other. 

The parameters $\lambda_l = 150$~eVnm$^{-2}$ and $\mu_l~=~109$~eVnm$^{-2}$ are the  first Lam{\'e} coefficient and shear modulus, respectively, of monolayer InSe. They are extracted (in a manner similar to that presented in Ref.~\onlinecite{PhysRevB.95.115409}) from the strain-dependence of the DFT energies of the primitive monolayer cells included in the MLIP training set. 

In Eq.~\eqref{eq:strain_cost} we neglect the energy cost of out-of-plane displacements of the InSe monolayers\cite{PhysRevLett.124.206101}, and assume that for each local lateral offset, $\mathbf{r}_0$, the optimal interlayer distance for an aligned bilayer with shift $\mathbf{r}_0$ is adopted, $d = d_{\textrm{min}}(\mathbf{r}_0)$. The relaxation procedure therefore requires an analytical approximation to the adhesion energy at the optimal interlayer distance for a given lateral shift
\begin{equation}
    \tilde{E}(\mathbf{r}_0) \equiv W(d=d_{\textrm{min}}(\mathbf{r}_0),\mathbf{r}_0).
\end{equation}
We expand $\tilde{E}$ as
\begin{equation}
    \tilde{E}(\mathbf{r}_0) =  \sum_n A_n e^{i\mathbf{G}_n\cdot \textbf{r}_0},
\end{equation}
where the Fourier coefficients $A_n$ are found by first minimising Eq. (\ref{eq:adhesion}) for a grid of $\mathbf{r}_0$ values, then taking a numerical Fourier transform of the results.
In a similar manner, one can express $d_{\textrm{min}}(\mathbf{r}_0)$ as
\begin{equation}
\label{eq:dmin_fourier}
    d_{\textrm{min}}(\mathbf{r}_0) = \sum_n D_n e^{i\mathbf{G}_n\cdot \mathbf{r}_0}.
\end{equation}
With the expansions and approximations set out above, we minimize Eq. \eqref{eq:competition} with respect to the Fourier components of the atomic displacement on reconstruction, $\mathbf{u}(\mathbf{r})$ (Eq. \eqref{eq:ur_expansion}).
\subsection{Comparison of continuum picture with atomistic relaxation}
\begin{figure}
    \centering
    \includegraphics[width = \linewidth]{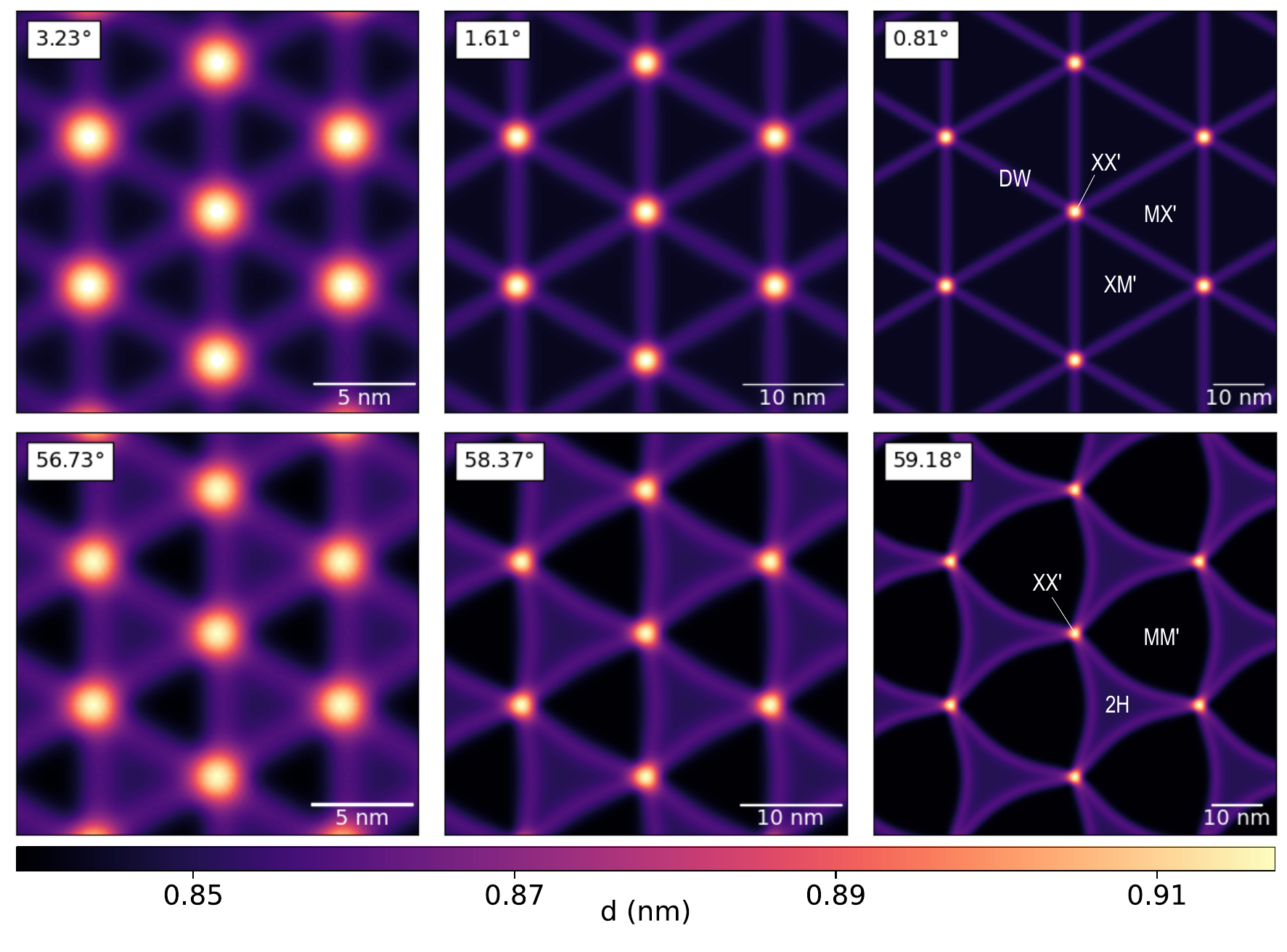}
    \caption{Interlayer distance maps from relaxation of twisted InSe using the continuum model.}
    \label{fig:cont_ild_maps}
\end{figure}
The results of the relaxation of twisted InSe bilayers using a continuum model approach are summarized in Fig.~\ref{fig:cont_ild_maps}. 
In terms of the in-plane shapes of the domains and domain walls, for P-stacking the continuum model reproduces the shapes of the domains and the domain wall thicknesses well as compared to the atomistic model.
However, for AP-stacking at angles closest to 60$^{\circ}$, the continuum model shows far less asymmetry in the shapes of the MM$'$ and 2H domains, with only a slight bending of the wall between them even at the smallest misalignment from AP twist. Therefore, whereas in the atomistic model, the 2H domain is already contracting to a node by 59.18$^{\circ}$, in the continuum model it is still largely triangular.

\begin{figure}
    \centering
    \includegraphics[width = \linewidth]{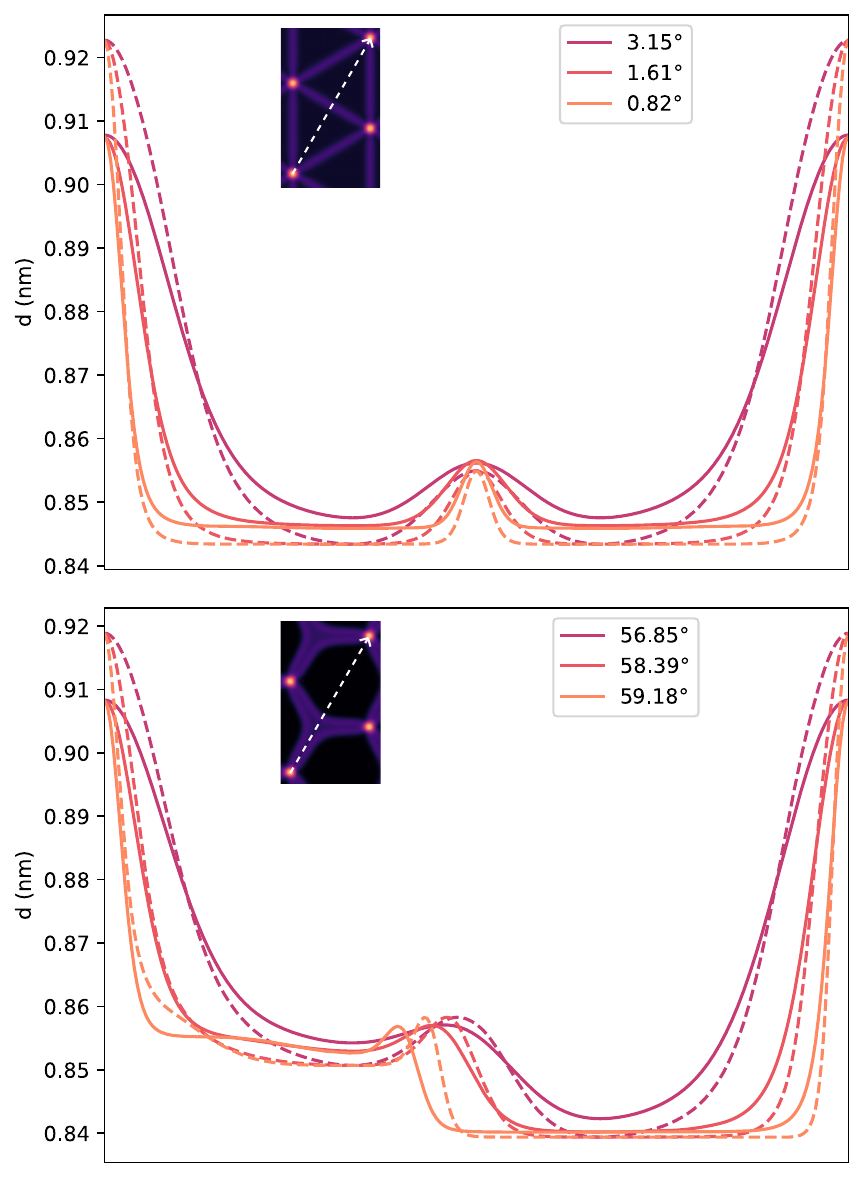}
    \caption{Comparison of interlayer distance along a path of length $\sqrt{3}L_M$ across the moiré supercell for twisted InSe bilayers relaxed using the trained MLIP (solid lines) and the continuum model (dashed lines).}
    \label{fig:cont_vs_mace_lines}
\end{figure}
As a further comparison between the continuum model and atomistic relaxation using the MLIP, we show plots of the interlayer distance along a line traversing the moiré supercell for both methods in Fig.~\ref{fig:cont_vs_mace_lines}. 
There are notable differences here for both the P- and AP-cases, arising from several sources.
The first is that the continuum model does not reproduce the predicted optimal interlayer distances for aligned primitive bilayer cells as well as the MLIP does. This means the predicted minimal interlayer distances in the MX$'$/XM$'$ stacked minima in the upper panel of Fig.~\ref{fig:cont_vs_mace_lines} differ between the two approaches, owing to the binding energy varying only slowly near the optimal interlayer distance in the continuum model. 
In terms of determining when domain structures form, these errors are only weakly significant, as the important factor is the differences between the binding energy minima for different stackings, not the interlayer distance at which they occur.
The second difference arises from the approximation in the continuum model that we can determine the interlayer distance in a twisted bilayer entirely from the local atomic registry.
The atomistic model, particularly around the XX' nodes, does not adhere to this approximation, as its interlayer distances around the nodes is significantly lower than the continuum model. 
For large twist angles, the layers remain rigid and the continuum model is, for obvious reasons, inadequate for determining interlayer distances across the whole supercell.

However, even for small twist angles, the small size of the features of the domain network (nodes and domain walls) means that we cannot neglect the energy cost of out-of-plane deformations.
For example, the XX$'$-stacked nodes have radii $\sim 1$~nm in the range $\theta\simeq 3^\circ$ to $1^\circ$, and this is no longer changing with reducing $\theta$ as the shape of the node is well converged by this point.
The change in interlayer distance on going from the centre to the edge of a node is $>$5\% of this value, so it is not correct to approximate them as locally flat.
The energy cost of bending the layers over such a short lengthscale therefore prevents the atomistically-relaxed interlayer distance at an XX$'$ stacked node from reaching a value as high as that predicted by the continuum model.
We observe a maximum interlayer distance of of 0.908~nm for the atomistic model versus 0.923~nm for the continuum model, in the P-stacked case.
Finally, the aforementioned failure of the continuum model to reproduce the shapes of the 2H/MM$'$ domains as predicted by the atomistic MLIP in the AP-stacked case can be seen from the relative displacement of the central peaked region (the domain wall) in the line cuts of the interlayer distance in the lower panel of Fig.~\ref{fig:cont_vs_mace_lines}.

\section{Conclusions}
\label{sec:conclusions}
In this work, we have trained and deployed a machine-learned interatomic potential to study atomistically the atomic reconstruction of twisted bilayer InSe, for cells as large as 39368 atoms.
We show how, as a consequence of its high flexibility, the structure of twisted InSe bilayers changes substantially from the rigid moiré case over a large range of twist angles and moiré supercell length scales. 
A comparison with an adapted continuum model shows that while a continuum approach reproduces several important features of twisted InSe bilayers, key differences remain due to the short length scale of some of the features in the relaxed domains, domain walls and nodes, highlighting the advantages of an atomistic approach in relaxing twisted bilayers of more flexible 2D materials. 

InSe has a bandstructure strongly sensitive to variations in interlayer hybridization on changes in stacking\cite{Kang_2020} and interlayer distance, to the application of strain\cite{Song2018, Zhao_2020}, and a strong piezoelectric response\cite{Li2015}. 
The properties of states hosted by twisted InSe bilayers can therefore be expected to depend strongly on the twist angle and on the details of the atomic structure within a moiré supercell, the details of which would merit further study.

The research data supporting this publication can
be accessed via the University of Warwick’s Research
Archive Portal \cite{wrap_archive}.

\begin{acknowledgments}
SJM acknowledges helpful discussions with V. I. Fal'ko, V. V. Enaldiev, and F. Ferreira. SJM and NDMH acknowledge funding from EPSRC grant number EP/V000136/1. AS acknowledges funding from the EPSRC CDT in Modelling of Heterogeneous Systems funded by EP/S022848/1. Computing facilities were provided by the Scientific Computing Research Technology Platform of the University of Warwick through the use of the High Performance Computing (HPC) cluster Avon, and the Sulis Tier 2 platforms at HPC Midlands+ funded by the Engineering and Physical Sciences Research Council (EPSRC), grant number EP/T022108/1. 
\end{acknowledgments}

\appendix

\end{document}